\documentclass[%
reprint,
 amsmath,amssymb,
 aps,
]{revtex4-1}

\usepackage{graphicx}
\usepackage{dcolumn}
\usepackage{bm}

\usepackage{soul}
\usepackage{cancel}
\usepackage{xspace}
\usepackage{lineno}
\usepackage{lineno,hyperref}
\usepackage{xcolor}

\newcommand{\pt}{\ensuremath{p_{\rm T}}\xspace}

\newcommand{\nmpi}{\ensuremath{N_{\rm mpi}}\xspace}

\newcommand{\VZERO}        {\rm{V0}\xspace}
\newcommand{\VZEROA}       {\rm{V0A}\xspace}
\newcommand{\VZEROC}       {\rm{V0C}\xspace}
\newcommand{\py}{PYTHIA\xspace}
\newcommand{\pPb}{p--Pb\xspace}

\begin{document}

\preprint{APS/123-QED}
\title{Unveiling the effects of multiple soft partonic interactions \\ in pp collisions at $\mathbf{\sqrt{{\textit s}}=13.6}$ TeV using charged-particle flattenicity}%
\author{Antonio Ortiz}
\email{antonio.ortiz@nucleares.unam.mx}
\affiliation{%
CERN, 1211 Geneva 23, Switzerland
}
\affiliation{%
Instituto de Ciencias Nucleares, UNAM,
 Apartado Postal 70-543, Ciudad de M\'exico 04510, M\'exico 
}

\author{Arvind Khuntia}
\affiliation{%
Faculty of Nuclear Sciences and Physical Engineering,
Czech Technical University in Prague, Brehova 7, 115 19, Prague,  Czech Republic 
}

\author{Omar V\'azquez Rueda}
\affiliation{%
Physics Department, University of Houston, 617 SR1 Building, Houston, TX 77204, USA
}

\author{Sushanta Tripathy}
\affiliation{%
INFN - sezione di Bologna, via Irnerio 46, 40126 Bologna BO, Italy
}

\author{Gyula Benc\'edi}
\affiliation{%
Wigner Research Centre for Physics, Hungary
}
\author{
Suraj Prasad}
\affiliation{
Department of Physics, Indian Institute of Technology Indore, Simrol, Indore 453552, India
}
\author{
Feng Fan}
\affiliation{
Central China Normal University, Wuhan, Hubei, 430079, China
}

\date{\today}

\begin{abstract}


Event classifiers based either on the charged-particle multiplicity or the event shape have been extensively used in proton-proton (pp) collisions by the ALICE collaboration at the LHC. The use of these tools became very instrumental since the observation of fluid-like behavior in high-multiplicity pp collisions. In particular, the study as a function of the charged-particle multiplicity registered in the forward V0 ALICE detector allowed for the discovery of strangeness enhancement in high-multiplicity pp collisions. However, one drawback of the multiplicity-based event classifiers is that requiring a high charged-particle multiplicity biases the sample towards hard processes like multi-jet final states. These biases make it difficult to perform jet-quenching searches in high-multiplicity pp collisions. In this context, the present paper explores the use of the new event classifier, flattenicity; which uses the multiplicity calculated in the forward pseudorapidity region.  To illustrate how this tool works, pp collisions at $\sqrt{s}=13.6$\,TeV simulated with PYTHIA~8 are explored. The sensitivity of flattencity to multi-partonic interactions as well as to the ``hardness'' of the collision are discussed. PYTHIA~8 predictions for the transverse momentum spectra of light- and heavy-flavored hadrons as a function of flattenicity are presented.      

\end{abstract}

\maketitle


\section{\label{sec:level1}Introduction}

High-multiplicity proton-proton (pp) and proton-led (\pPb) collisions, hereinafter referred to as small-collision systems, at ultra-relativistic energies have unveiled similarities with heavy-ion collisions~\cite{ALICE:2016fzo,CMS:2016fnw,Busza:2018rrf}. The origin of these effects in small-collision systems is still an open question in the heavy-ion community, where there is no evidence of jet quenching in such collisions~\cite{Nagle:2018nvi}. According to event generators like PYTHIA~8~\cite{Sjostrand:2007gs}, a high-activity at midpseudorapidity (high-multiplicity pp collisions) can be originated by two mechanisms. Either by several semi-hard parton-parton interactions occurring within the same pp collision, a phenomenon that is known as multi-partonic interactions (MPI)~\cite{Sjostrand:1987su}, or by multi-jet final states~\cite{Ortiz:2016kpz}. Since the goal of the study is to establish whether a small drop of strongly-interacting quark-gluon plasma (sQGP) is formed in small collision systems, one has to isolate high-multiplicity pp collisions originated by soft partonic processes. 

Multi-partonic interactions offer an alternative approach to explain the observed fluid-like phenomena in high-multiplicity pp collisions. For instance, color reconnection (CR) can mimic radial flow patterns in pp collisions with a large number of MPI (\nmpi)~\cite{Ortiz:2013yxa}.  Models based on the QCD theory of MPI can partially explain collectivity from interference effects in hadronic collisions with \nmpi parton-parton scatterings~\cite{Blok:2017pui,Blok:2018xes}.  PYTHIA~8 with the rope hadronization model~\cite{Bierlich:2015rha}, which assumes the formation of ropes due to overlapping of strings in a high-multiplicity environment (high \nmpi), describes the strangeness enhancement~\cite{Nayak:2018xip}. Regarding the phenomena at large transverse momentum ($\textit{p}_{T}$), the model also produces some features that are present in data from heavy-ion collisions~\cite{Mishra:2018pio,Jacobs:2020ptj,Ortiz:2020dph}. 

\begin{figure*}[t]
\includegraphics[width=0.46\textwidth]{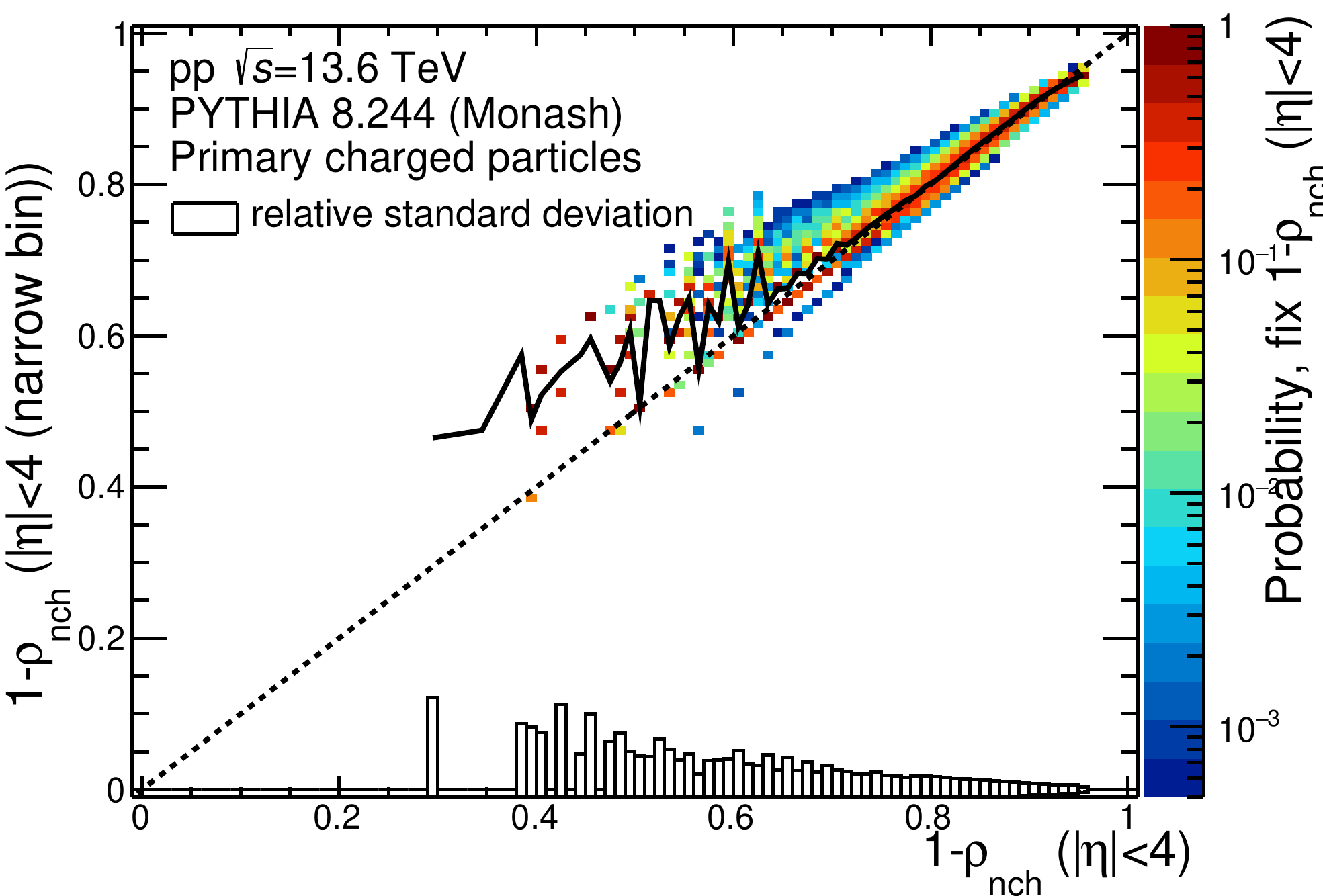}
\includegraphics[width=0.46\textwidth]{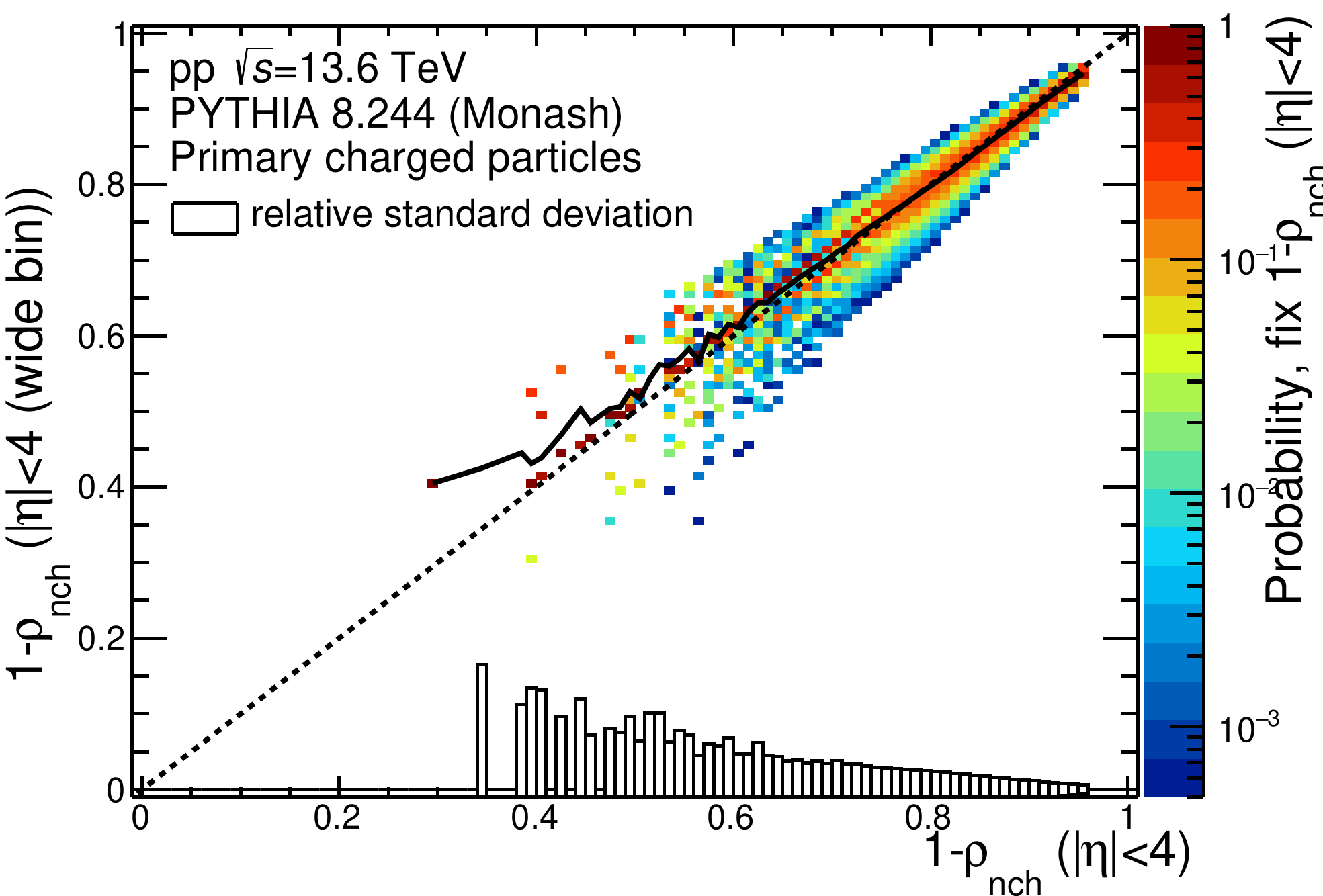}
\caption{One minus flattenicity calculated using a narrow (wide) binning in $\eta-\varphi$ (read the text for more details) as a function of the reference one minus flattenicity. Results for pp collisions at $\sqrt{s}=13.6$\, TeV simulated with \py are displayed. The boxes around zero indicate the relative standard deviation as a function of the reference flattenicity.}
\label{fig1}  
\end{figure*}

One of the most common event classifiers in ALICE is based on the measurement of the charged-particle multiplicity in a different pseudorapidity interval to that where the observable of interest is measured. Different charged-particle multiplicity classes are defined based on the total charge deposited in the \VZERO detector, hereinafter referred to as V0M multiplicity. The \VZERO consists of two arrays of scintillating tiles placed on each side of the interaction point covering the full azimuthal acceptance and the pseudorapidity intervals of 2.8~$<\eta<$~5.1 (\VZEROA) and -3.7~$<\eta<$~-1.7 (\VZEROC). With this approach, the strangeness enhancement was discovered in pp collisions~\cite{ALICE:2017jyt}. Alternative studies that use event shape observables like transverse sphericity~\cite{ALICE:2012cor}, transverse spherocity~\cite{ALICE:2019dfi,Ortiz:2015ttf}, and the relative transverse activity classifier~\cite{Martin:2016igp, Ortiz:2017jaz} have tried to isolate the soft particle production. However, in all studies reported so far, the presence of undesired bias makes it difficult to interpret the results. For example, the attempt to search for jet quenching effects in pp collisions has not been successful~\cite{Krizek:2021sla,ALICE:2022qxg}. Although a significant broadening is observed in the acoplanarity distribution of high-multiplicity events, consistent with jet quenching, the same effect is present in models that do not include the effects of a medium. The simulations suggest that the enhanced acoplanarity results from the bias induced by the high-multiplicity selection towards multi-jet final states.     

In this paper, flattenicity is explored, it is intended to be sensitive to soft multi-partonic interactions, and it is accessible to experiments at the LHC. All results presented in this paper correspond to pp collisions at $\sqrt{s}=13.6$\,TeV simulated with PYTHIA~8.307~\cite{Bierlich:2022pfr}. The PYTHIA~8 tunes used in the paper are described in section~\ref{sec:mc}. In section~\ref{sec:analysis} flattenicity is defined and tested under variations in the segmentation used to calculate it. In section~\ref{sec:V0vsFlat} a comparison between the widely used V0M estimator and flattenicity is discussed. Section~\ref{sec:results} presents studies of light- and heavy-flavor production as a function of flattenicity. Finally, we summarise our results with an outlook for the upcoming ALICE run~3 and 4 measurements in section~\ref{sec:conclusions}.

\section{\label{sec:mc}The PYTHIA~8 event generator: Monash vs CR mode 2}

PYTHIA~8 simulations with the models Monash and the QCD-based color reconnection mode 2 (CR2) are used in the present study. The main features of the models are briefly described in this section.

PYTHIA~8~\cite{Sjostrand:2014zea} is one of the most widely used Monte Carlo event generators for high-energy collider physics with particular emphasis on physics related to small collision systems such as pp collisions. It is a parton-based microscopic event generator, where the main event of a pp collision is represented with hard parton scatterings via $2\rightarrow2$ matrix elements defined at leading order. It is then complemented by the leading-logarithm approximation of parton showers that includes initial- and final-state radiation. The underlying event is formed by particles originating from MPI as well as from beam remnants. The hadronization from partons is performed using the Lund string fragmentation model~\cite{Andersson:1983ia}. In the color reconnection (CR) picture~\cite{Argyropoulos:2014zoa}, the strings between partons can be rearranged in a way that the total string length is reduced; by which the total charged-particle multiplicity of the event is also reduced. The Monash 2013 tune of PYTHIA~8~\cite{Skands:2014pea}, created for a better description of minimum-bias and underlying-event observables in pp collisions at the LHC energies, includes the MPI-based color-reconnection scheme. In this scheme, the color flow relies on the parton shower-like color configuration of the beam remnants, and partons are classified based on their origin from the MPI system. However, the color rules of QCD in the beam remnant are not considered in the MPI-based color reconnection scheme. Recently, a newer QCD-based CR scheme is introduced, which encompasses the minimization of the string length as well as the color rules from QCD~\cite{Bierlich:2015rha}. This new CR scheme introduces several tuneable parameters and it has been recently shown that the baryon-to-meson ratio from pp collisions at the LHC is better explained by a set of parameters, referred to as CR2~\cite{Bierlich:2015rha,ALICE:2021rzj,ALICE:2020wla}.

\section{\label{sec:analysis}Flattenicity}

Inspired by the recently introduced flattenicity~\cite{Ortiz:2022zqr}, that is proposed as a new observable to be measured in the next-generation heavy-ion experiment at CERN (ALICE 3) in the LHC Run~5~\cite{ALICE:2803563}. The present work explores the feasibility of flattenicity measurement using the existing detectors of ALICE~\cite{ALICE:2014sbx}. For the definition of flattenicity, the $\eta-\varphi$ phase space was divided into $\rm N_{cell}=80$ elementary cells. Given the expected tracking capabilities of ALICE 3, charged particles within $|\eta|<4$ and $p_{\rm T}>0.15$\,GeV/$c$ were considered in the calculation of flattenicity. In the cell $i$,  the average transverse momentum was calculated ($p_{\rm T}^{{\rm cell},i}$). Event-by-event, the relative standard deviation defines flattenicity as follows:

\begin{equation}
\rho =\frac{\sqrt{\sum_{i}{(p_{\rm T}^{{\rm cell},i} - \langle p_{\rm T}^{\rm cell} \rangle)^{2}}/{\rm N_{cell}}}}{\langle p_{\rm T}^{\rm cell} \rangle},
\end{equation} 

Events with jet signals on top of the underlying event are expected to have a large spread in the $p_{\rm T}^{{\rm cell},i}$ values, the opposite is expected in the case in which particles with lower momenta would be isotropically distributed. The aim of this work is to redefine flattenicity in such a way that it can be measured by the main four experiments at the LHC using the existing data. Therefore, charged-particle multiplicity is used instead of the average transverse momentum per cell. In order to guarantee values of flattenicity between 0 and 1, like the standard event shapes (see e.g.~\cite{Ortiz:2017jho}), the event shape is defined as follows:

\begin{equation}
\rho_{\rm nch}=\frac{\sqrt{\sum_{i}{(N_{\rm ch}^{{\rm cell},i} - \langle N_{\rm ch}^{\rm cell} \rangle)^{2}}/{\rm N_{cell}^{2}}}}{\langle N_{\rm ch}^{\rm cell} \rangle},
\end{equation} 
where, $N_{\rm ch}^{{\rm cell},i}$ is the average multiplicity in the elementary cell $i$ and $\langle N_{\rm ch}^{\rm cell} \rangle$ is the average of $N_{\rm ch}^{{\rm cell},i}$ in the event. Flattenicity is calculated in the pseudorapidity intervals specified along the paper and using primary charged particles with $p_{\rm T}>0$. The additional factor $1/\sqrt{\rm N_{cell}}$ guarantees flattenicity to be smaller than unity. Moreover, in order to have a similar meaning of the limits of the new event shape to those used so far (e.g. spherocity), this paper reports results as a function of $1-\rho_{\rm nch}$. In such a way that events with $1-\rho_{\rm nch}\rightarrow1$ are associated with the isotropic topology, whereas those with  $1-\rho_{\rm nch}\rightarrow0$ are associated with jet topologies.

In the following studies, the so-called reference flattenicity was calculated considering 16 bins in pseudorapidity (bin size 0.5) and 8 bins in $\varphi$ (bin size $2\pi/8\approx0.79$\,rad). In order to check the stability of flattenicity with the change of the cell size, a comparison between the reference flattenicity and two variations was cross-checked. The wide cell case consists of 8 and 6 equal-sized intervals in pseudorapidity ($-4<\eta<4$) and $\varphi$ ($0<\varphi<2\pi$), respectively. For the narrow cell case, 32 and 25 equal-sized intervals were considered for $\eta$ and $\varphi$, respectively. Figure~\ref{fig1} shows the correlation between the flattenicity obtained using a narrow (or wide) binning in $\eta-\varphi$ and the reference flattenicity. Results for non-diffractive pp collisions at $\sqrt{s}=13.6$\,TeV simulated with PYTHIA~8 tune Monash~\cite{Skands:2014pea} are shown. The relative spread is within 5\%, and the average values are consistent within a few percentages. This result shows the stability of flattenicity against variations in the size of the cells. This makes feasible the flattenicity measurement  in experiments like ALICE, which would rely on detectors with a given segmentation in $\eta$ and $\varphi$.     

\begin{figure}[t]
\includegraphics[width=0.46\textwidth]{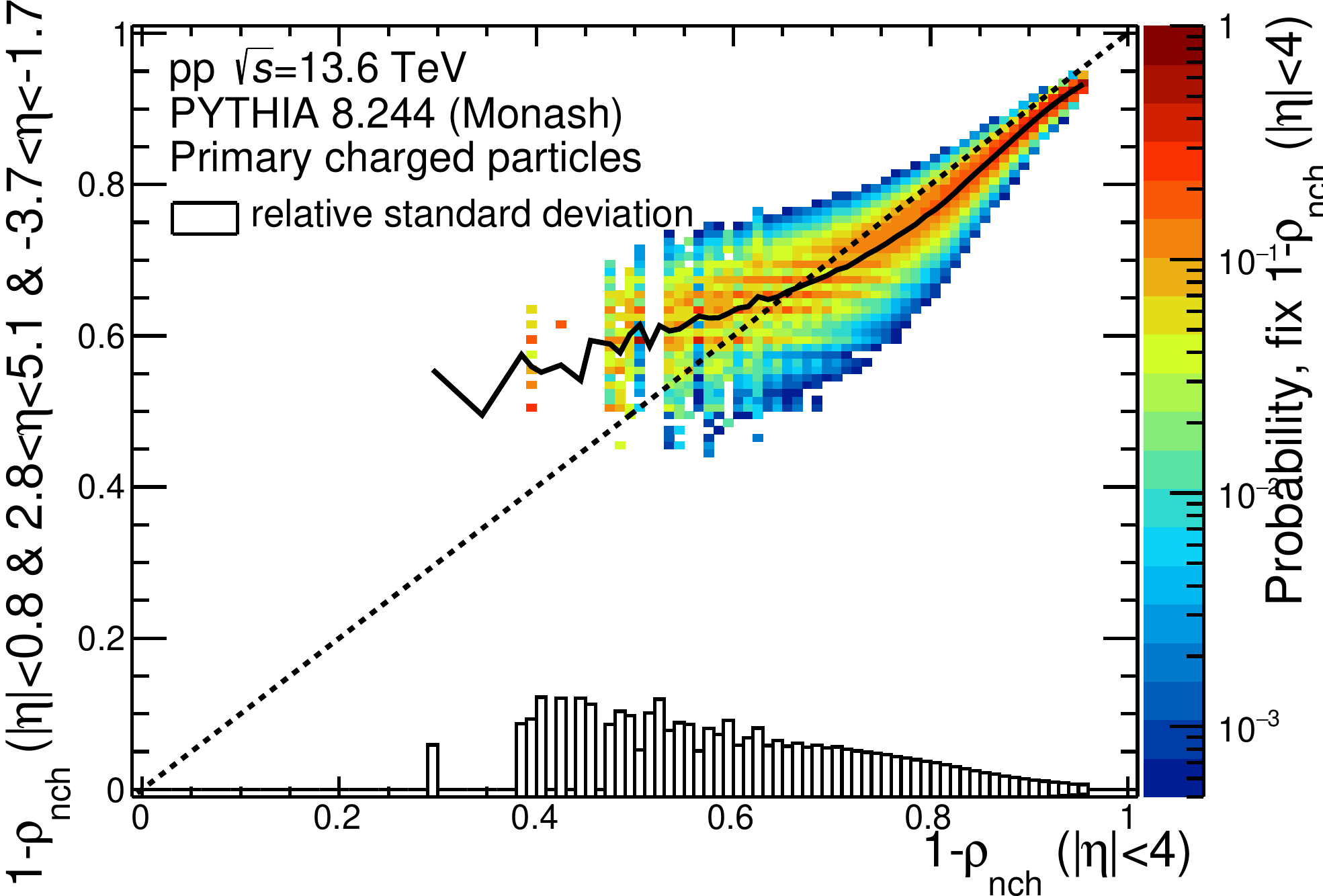}
\caption{One minus flattenicity calculated using the experimental accessible segmentation of ALICE  ($-3.7<\eta<-1.7$, $|\eta|<0.8$ and $2.8<\eta<5.1$) as a function of the reference one minus flattenicity. Results for pp collisions at $\sqrt{s}=13.6$\,TeV simulated with \py are displayed. The boxes around zero indicate the relative standard deviation as a function of the reference flattenicity.}
\label{fig2}  
\end{figure}

Regarding the current capabilities of ALICE, several results as a function of the V0M charged-particle multiplicity have been reported. Each V0 subdetector is segmented into four rings covering the pseudorapidity intervals listed in Table~\ref{tab:1}. Each ring is divided into eight equal-sized intervals in the azimuth. This
yields 64 sectors and the multiplicity in each sector can be used to calculate flattenicity. In addition, the charged-particle multiplicity at mid-pseudorapidity ($|\eta|<0.8$) which is measured using the TPC of ALICE can be used to constrain the flattenicity of the events. Therefore, a grid defined by the 64 \VZERO sectors can be complemented with an additional grid within $|\eta|<0.8$ and $0<\varphi<2\pi$ formed by 32 equal-sized cells ($0.5 \times 2\pi/8$\,rad)~\cite{ALICE:2013axi}. Figure~\ref{fig2} shows the correlation between the flattenicity obtained using such a segmentation ($-3.7<\eta<-1.7$, $|\eta|<0.8$ and $2.8<\eta<5.1$) and the reference flattenicity ($|\eta|<4$). The experimentally accessible flattenicity is shifted by up to 5\% with respect to the reference flattenicity. For a fixed value of reference flattenicity, the relative standard deviation goes from 10\% to less than 1\% from low to high  $1-\rho_{\rm nch}$ values.    

\begin{table} [!hpt]
                \centering 
                \caption{Pseudorapidity intervals covered by the different rings of the \VZERO detector of ALICE.
                 \label{tab:1}}
                \begin{tabular}{|c |c | c |}
                \hline
                \textbf{Ring}  & \textbf{\VZEROC} & \textbf{\VZEROA} \\ 
                \hline
                1       & $-3.7<\eta<-3.2$  & $4.5<\eta<5.1$  \\
                2       & $-3.2<\eta<-2.7$  & $3.9<\eta<4.5$  \\
                3       & $-2.7<\eta<-2.2$  &  $3.4<\eta<3.9$ \\
                4       & $-2.2<\eta<-1.7$  &  $2.8<\eta<3.4$ \\
                \hline
                \end{tabular}
\end{table}

\begin{figure*}[hbt!]
\includegraphics[width=0.46\textwidth]{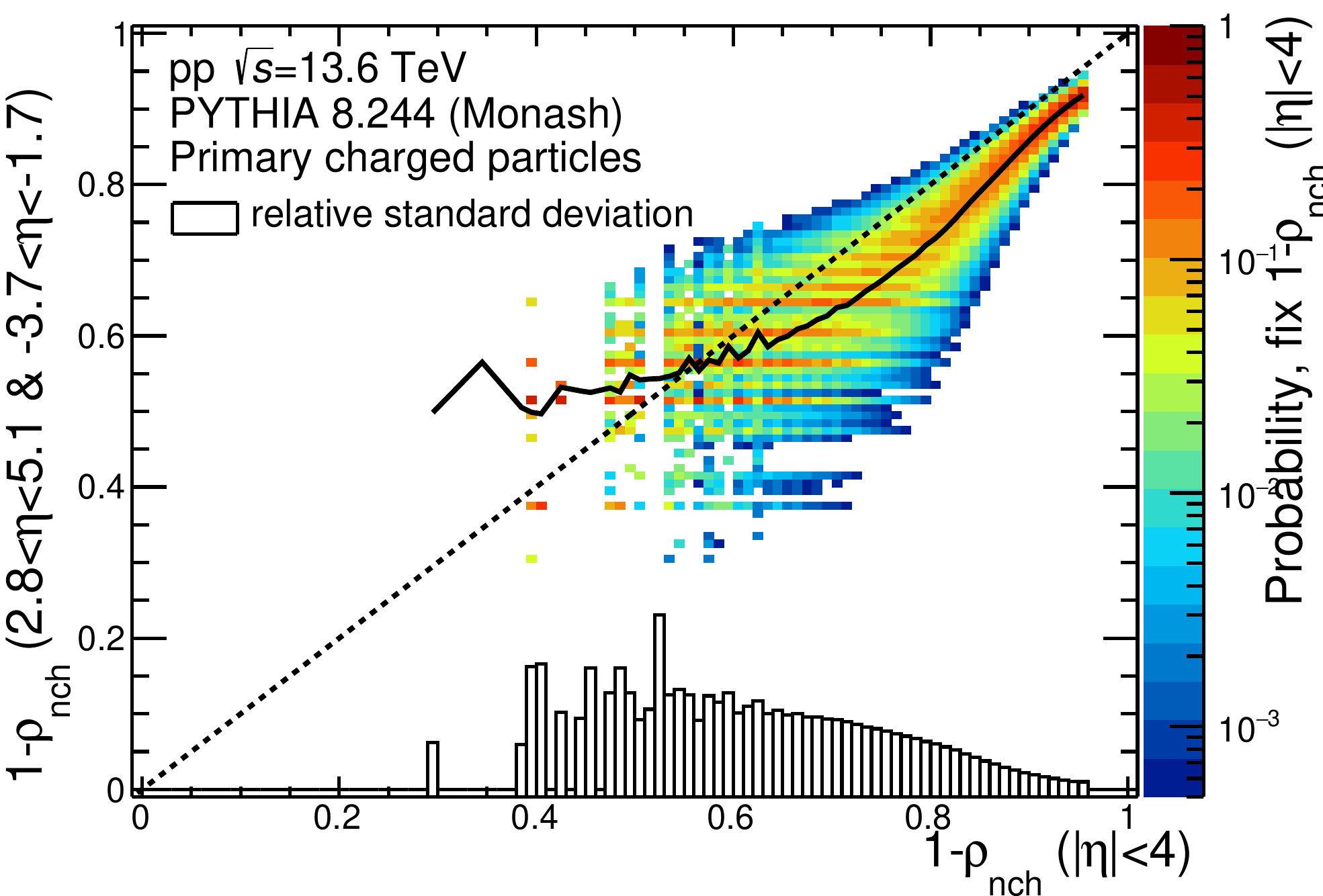}
\includegraphics[width=0.46\textwidth]{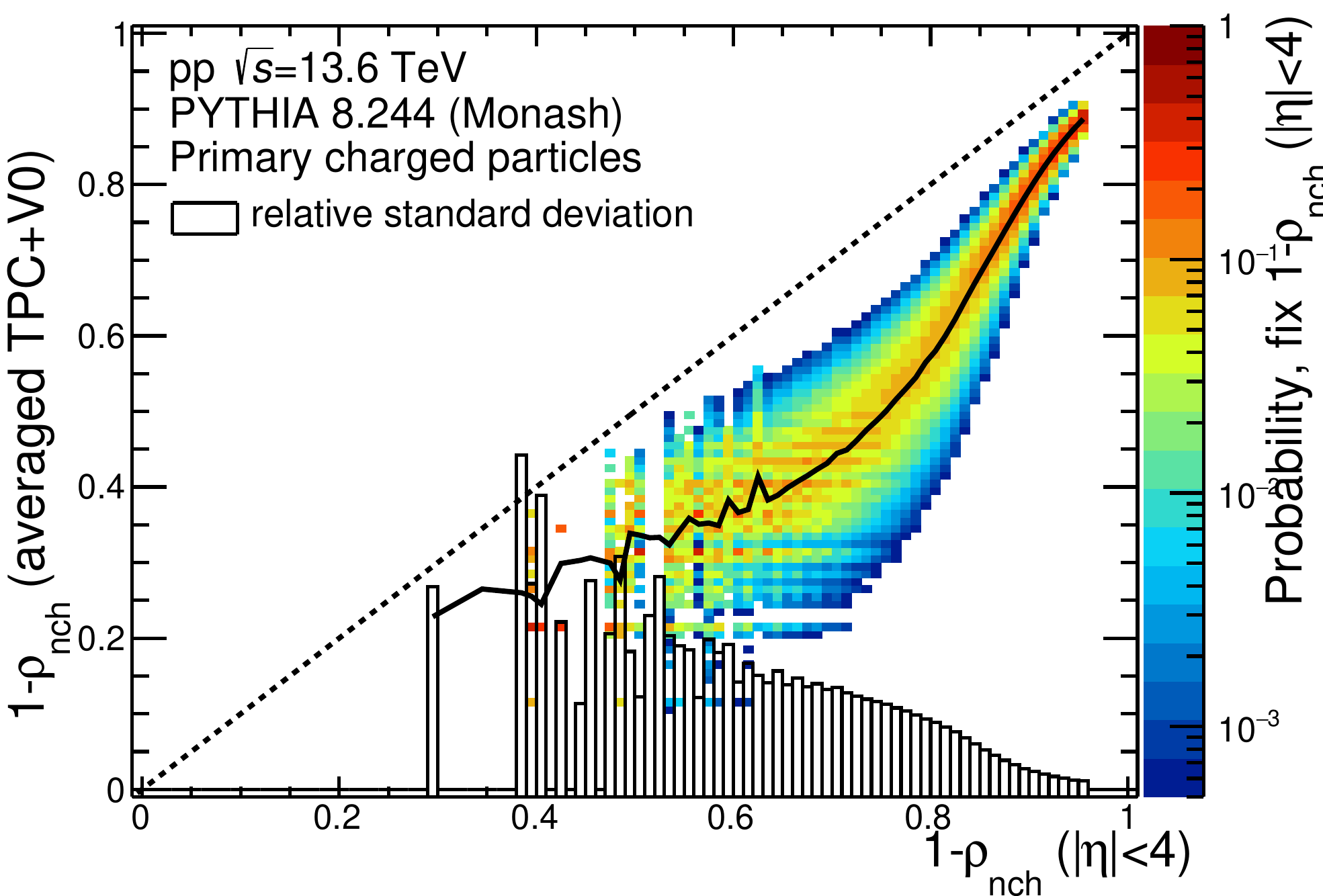}
\caption{Left (right): one minus flattenicity calculated in the pseudorapidity interval covered by the ALICE \VZERO (TPC) detector (read the text for more details) as a function of one minus reference flattenicity. Results for pp collisions at $\sqrt{s}=13.6$\,TeV simulated with \py are displayed. The boxes around zero indicate the relative standard deviation as a function of the reference flattenicity.}
\label{fig3}  
\end{figure*}

\begin{figure*}[hbt!]
\includegraphics[width=0.46\textwidth]{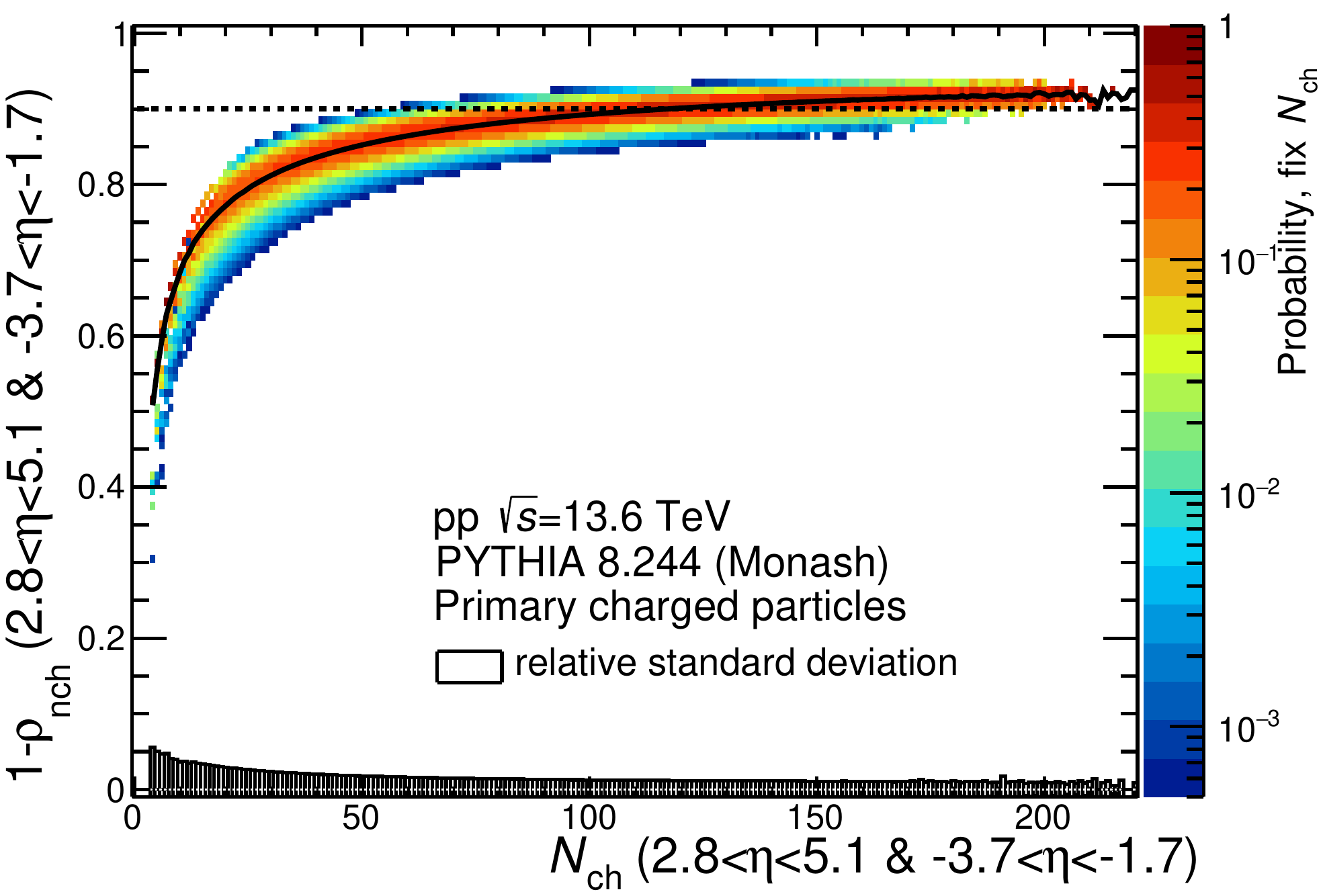}
\includegraphics[width=0.46\textwidth]{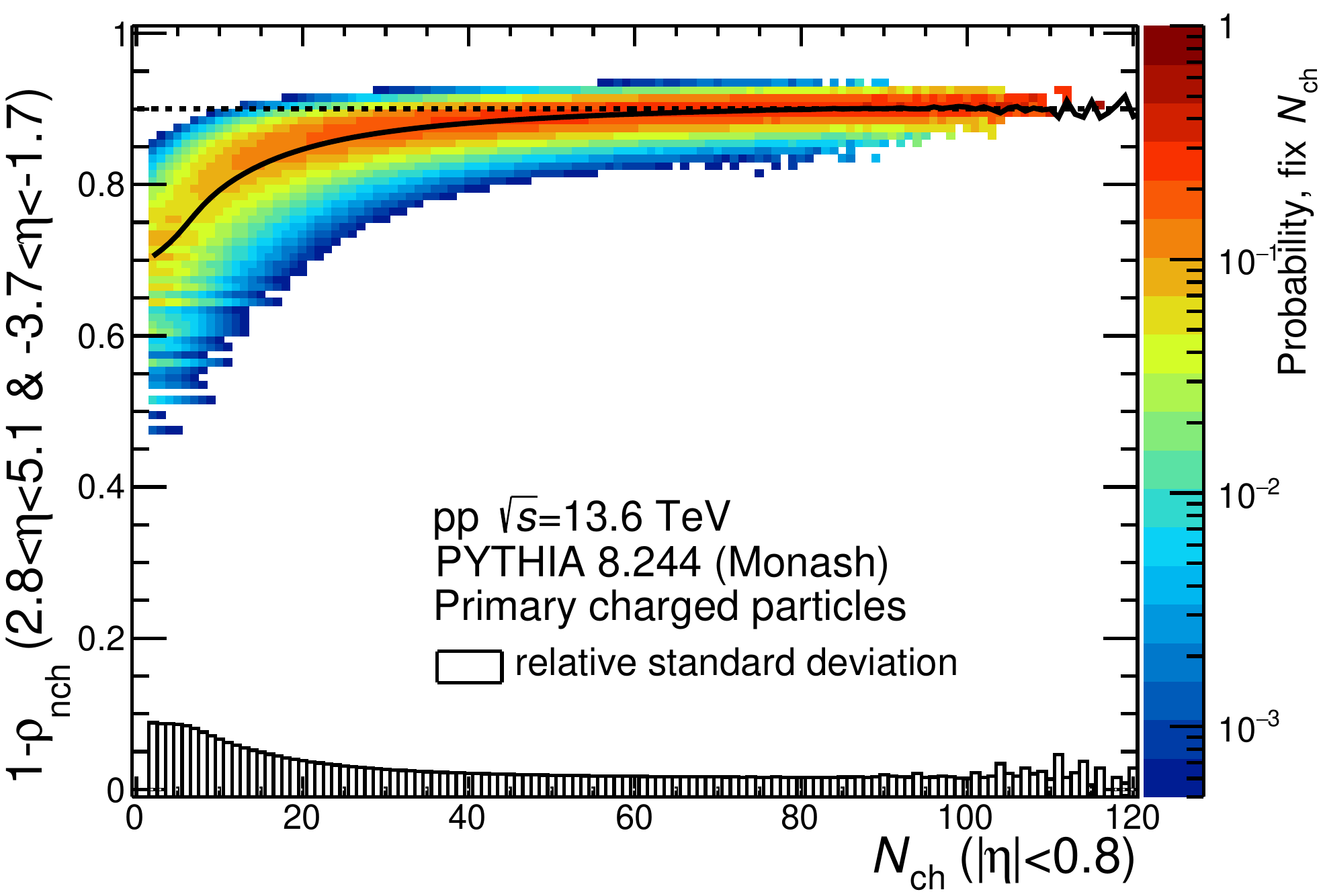}
\caption{One minus flattenicity calculated in the pseudorapidity region covered by the \VZERO detector. The event shape is plotted as a function of the charged-particle multiplicity in the same pseudorapidity interval (left) and at midpseudorapidity (right). The width of the distribution is shown as boxes around zero.}
\label{fig4}  
\end{figure*}

The measurement of flattenicity using the segmentation previously defined can be a bit tricky, because the \VZERO detector provides charge amplitude. Of course, one can develop a strategy to combine the information from the two detectors. However, it would be straightforward to determine flattenicity in a detector-independent way. To this end, Fig.~\ref{fig3} explores the correlation between flattenicity calculated in the pseudorapoidity region covered by the \VZERO detector and the reference flattenicity. The figure also displays the situation in which flattenicity is calculated at midpseudorapidity and \VZERO. The combined flattenicity is given by the average of the flattenicity values obtained in each case and it is termed as average TPC$+$\VZERO. The average flattenicity values obtained in the acceptance of the \VZERO (\VZERO$+$TPC) detector are shifted down by up to 10\% (40\%) with respect to the reference flattenicity values. On the other hand, for low $1-\rho_{\rm nch}$ values obtained in the acceptance of the \VZERO (\VZERO$+$TPC) detector the relative standard deviation is around 10\% (40\%) and decreases to less than 1\% at high $1-\rho_{\rm nch}$. The effect is explained as due to a bias towards hard pp collisions when the event activity is calculated at midpseudorapidity. The result suggests that a measurement of flattenicity in the \VZERO acceptance would be the best to enhance the sensitivity to the global event shape. This is the approach that is followed in the present work.

\section{\label{sec:V0vsFlat}V0M vs flattenicity}

Given the definition of flattenicity, the event classifier is expected to be strongly multiplicity dependent. This means that the limit $1-\rho_{\rm nch}\rightarrow1$ can be easily reached by high-multiplicity events, whereas for jet-like events,  $1-\rho_{\rm nch}\rightarrow0$ would be easily reached by low-multiplicity events.

\begin{figure*}[!ht]
\begin{center}
\includegraphics[width=0.46\textwidth]{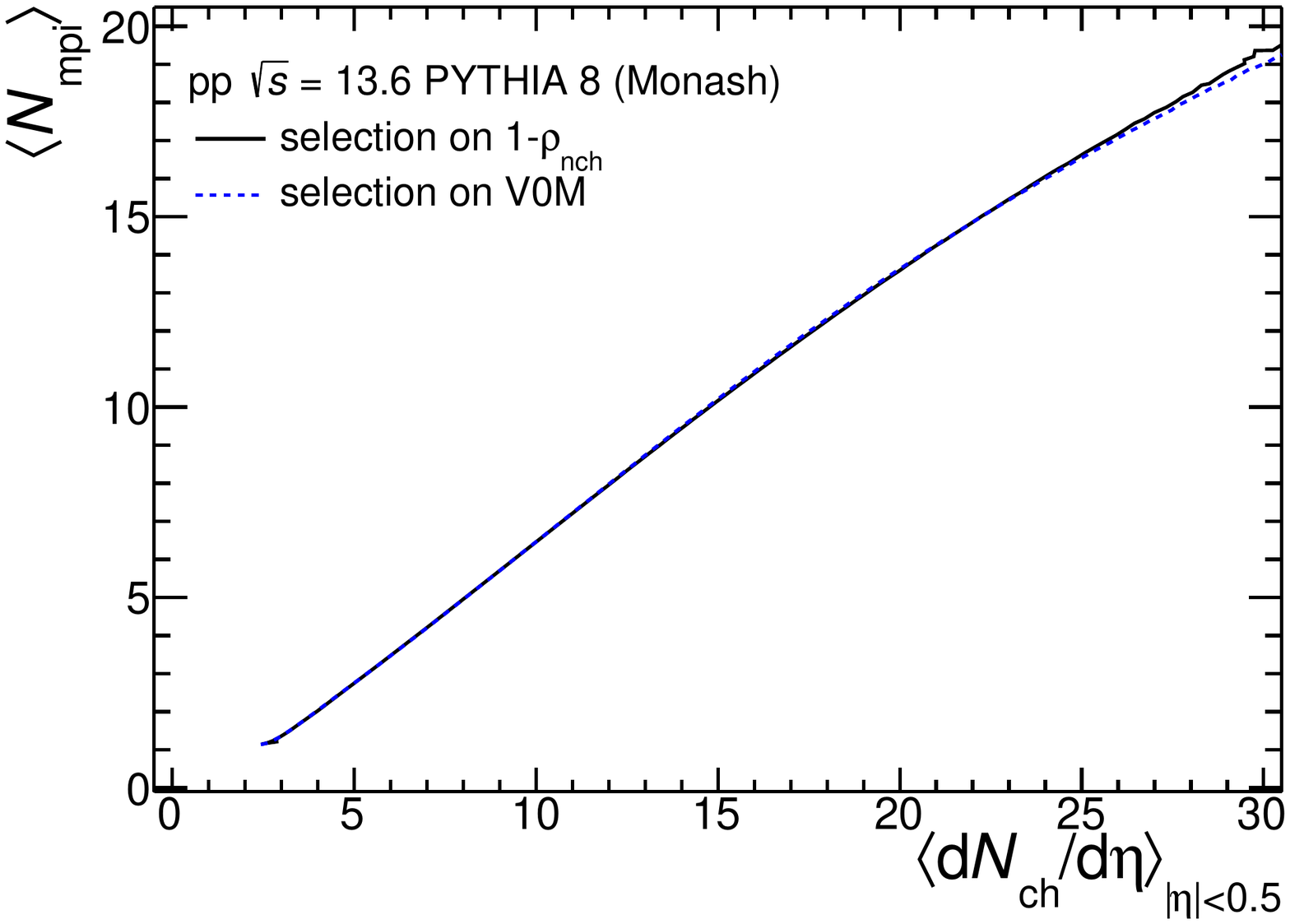}
\includegraphics[width=0.46\textwidth]{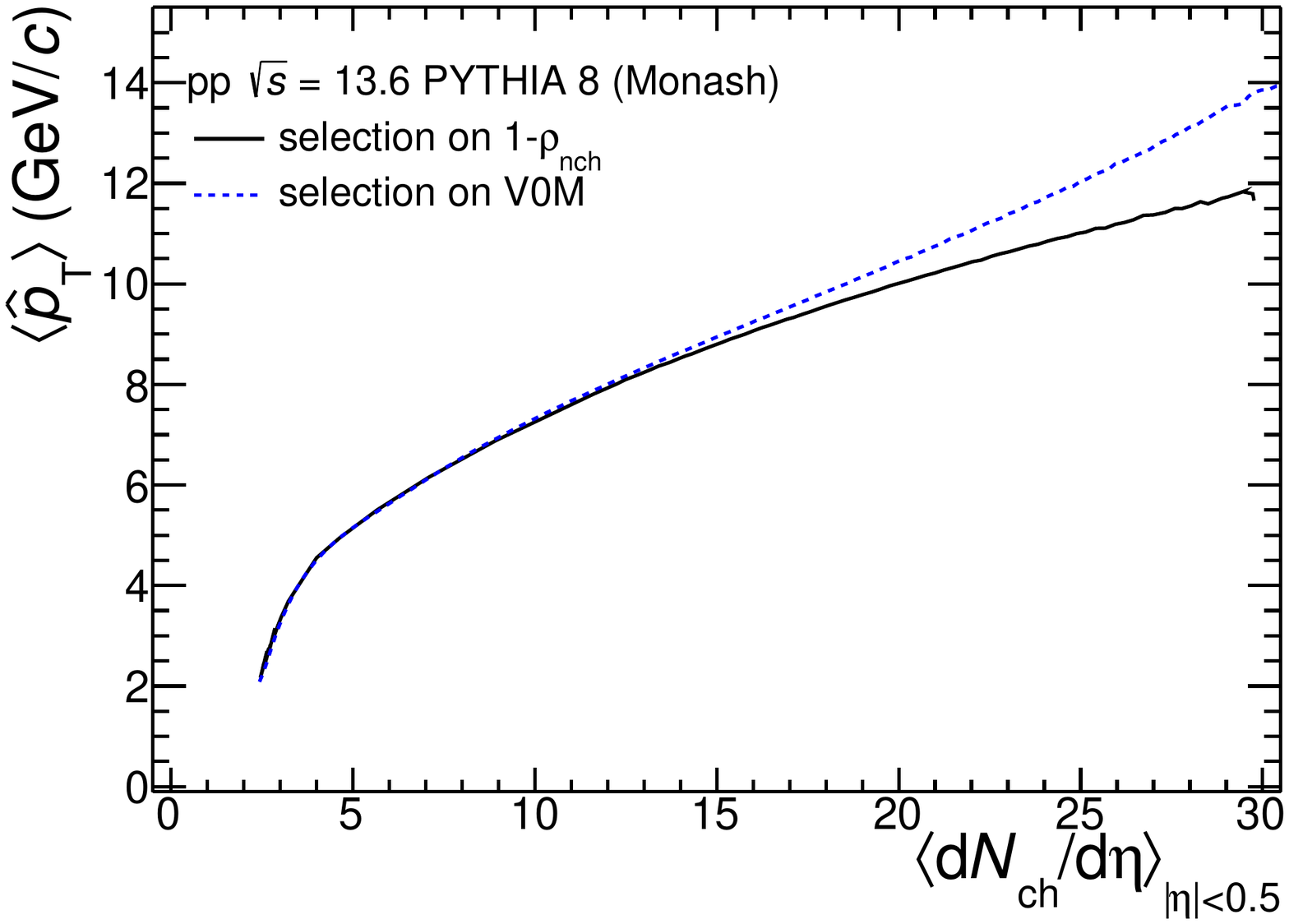}
\caption{The average number of multi-partonic interactions (transverse momentum of the main partonic scattering) as a function of the charged-particle multiplicity density at midpsudorapidity, $|\eta| <$  0.5, is shown in the left (right) hand side panel. Results for pp collisions at $\sqrt{s}=13.6$\,TeV simulated with PYTHIA~8 tune Monash are displayed. The solid line indicates the correlation when the event selection is done in $1-\rho_{\rm nch}$ classes, whereas, the dotted line indicate the correlation when the event classification is performed as a function of the V0M estimator.}
\label{fig5}  
\end{center}
\end{figure*}

This feature of flattenicity is illustrated in Fig.~\ref{fig4} where the correlation between $1-\rho_{\rm nch}$ and multiplicity is shown for pp collisions simulated with PYTHIA~8 tune Monash. As in the previous sections, for a fixed multiplicity value the relative standard deviation is displayed. If the multiplicity is determined in pseudorapidity interval covered by the \VZERO detector, $1-\rho_{\rm nch}$ exhibits a rise even at high multiplicities, where it reaches values above 0.9. The dependence with multiplicity at midpseudorapidity shows saturation at $1-\rho_{\rm nch}=0.9$ which is reached for intermediate multiplicities (${\rm d}N_{\rm ch}/{\rm d}\eta \approx 38$). However, in this case the distribution is slightly wider.  The effects can be factorized by performing an analysis both as a function of multiplicity and flattenicity. 

Different event classes will be studied based on percentiles of either multiplicity or flattenicity both of them calculated in the pseudorapidity region covered by the \VZERO detector of ALICE. Table~\ref{tab:2} shows the average charged-particle multiplicity density at midpseudorapidity for different event classes defined with flattenicity or V0M multiplicity classes. For similar percentiles, the average multiplicity values are very close to each other. However, as it will be shown later, the characteristics of the events are very different. 

\begin{figure*}[!ht]
\begin{center}
\includegraphics[width=0.46\textwidth]{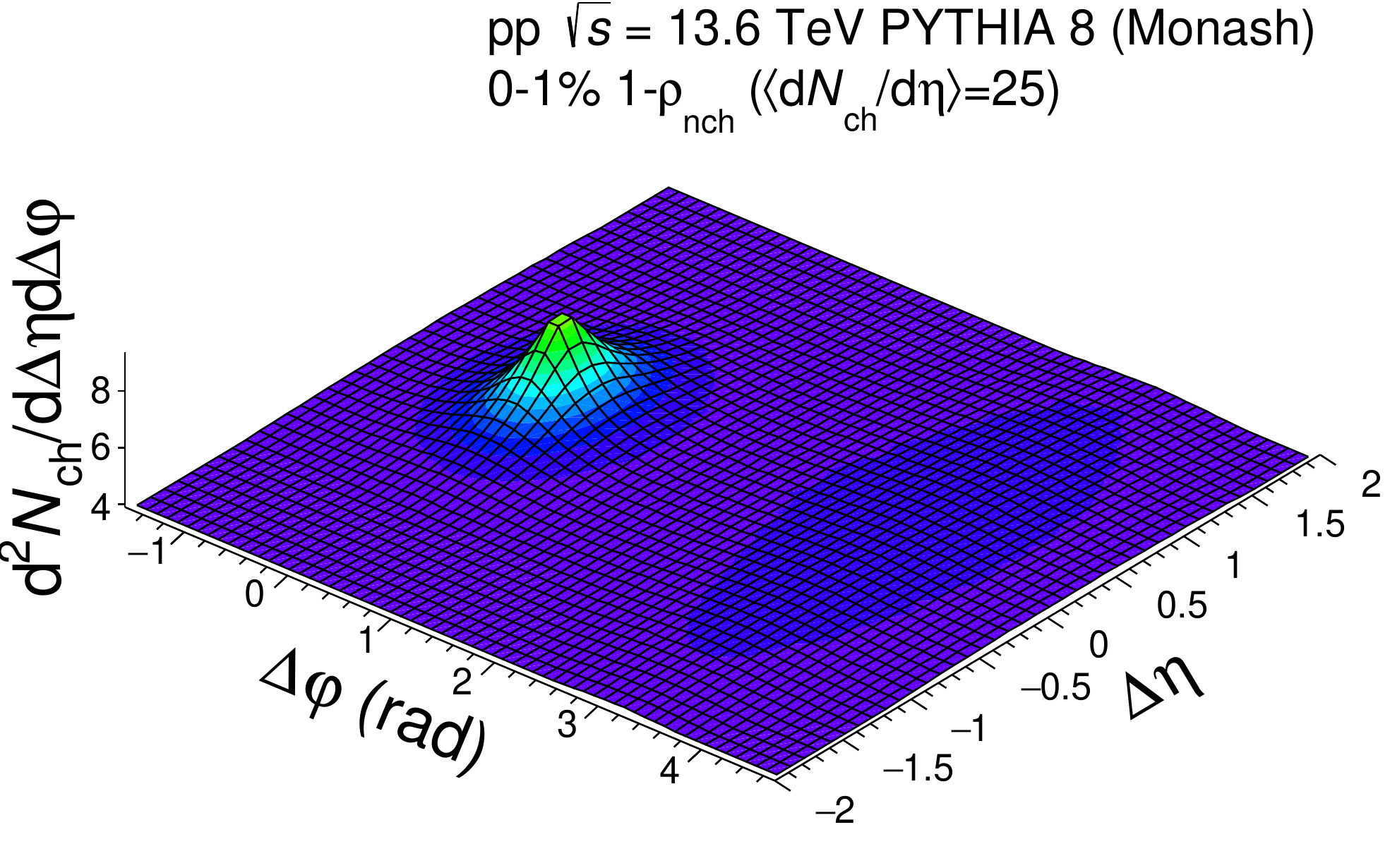}
\includegraphics[width=0.46\textwidth]{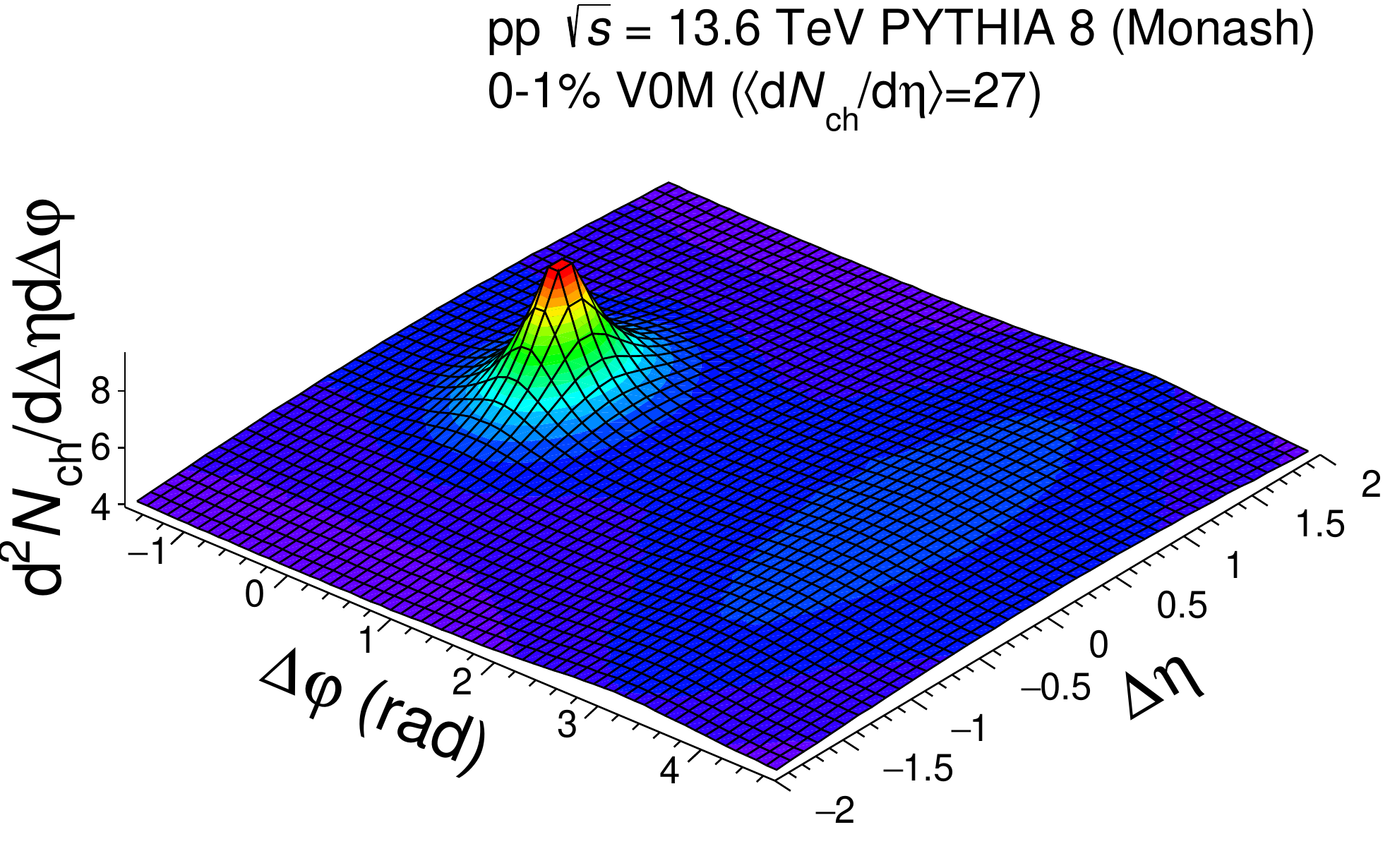}
\caption{Angular correlations between associated charged particles and the leading charged particle. The charged-particle yield is presented as a function of the angular separation $\Delta\varphi$ and $\Delta\eta$ for pp collisions at $\sqrt{s}=13.6$\,TeV simulated with PYTHIA~8 tune Monash. The correlation for the 0-1\% $1-\rho_{\rm nch}$ (V0M) event class is shown in the left (rigth) panel.}
\label{fig6}  
\end{center}
\end{figure*}

\begin{table} [!hpt]
                \centering 
                \caption{Average charged-particle multiplicity density ($\langle {\rm d}N_{\rm ch}/{\rm d}\eta\rangle$) at midpseudorapidity ($|\eta|<0.8$) for different percentile classes defined using flattenicity and V0M.
                 \label{tab:2}}
                \begin{tabular}{|c |c | c |}
                \hline
                \textbf{Event class}  & $\mathbf{1-\rho_{\rm nch}}$ & \textbf{V0M} \\ 
                \hline
                0-1\%       & 25.0   & 27.1  \\
                1-5\%       & 22.9  & 23.0  \\
                5-10\%       & 18.4  & 18.7  \\
                10-20\%       & 15.6  & 15.3  \\
                \hline
                \end{tabular}
\end{table}

Table~\ref{tab:2} hints that one can select pp collisions with similar charged-particle multiplicity at midpseudorapidity but originated from different processes. The left-hand side plot shown in Fig.~\ref{fig5} displays the correlation between the average number of multi-partonic interactions and the charged-particle multiplicity density at $|\eta|<0.5$. A comparison of the correlation obtained either using a even selection based on flattenicity or V0M multiplicity is displayed. In both cases, the average number of multi-partonic interactions  increases with the increase of the event activity estimator (V0M multiplicity or $1-\rho_{\rm nch}$). Moreover, both selections give a very similar linear correlation, however, slightly higher $\langle N_{\rm mpi} \rangle$ values are observed when the event selection is done with flattenicity. In order to study the ``hardness'' of the samples, the right-hand side of Fig.~\ref{fig5}~shows the average transverse momentum of the main parton-parton scattering (${\hat p}_{\rm T}$) as a function of the charged-particle multiplicity density. For $\langle {\rm d}N_{\rm ch}/{\rm d}\eta\rangle<15$ both estimators give the same result. In both cases a  fast increase of $\langle {\hat p}_{\rm T} \rangle$ is observed for multiplicity densities below 5, this behavior is followed by a reduction in the slope of  $\langle {\hat p}_{\rm T} \rangle$ as a function of the multiplicity density. However, for higher multiplicities a clear deviation is observed between the two event classifiers. The selection in terms of V0M gives a steeper rise of $\langle {\hat p}_{\rm T}\rangle$ with the charged-particle density than that observed for the selection based on flattenicity. At $\langle {\rm d}N_{\rm ch}/{\rm d}\eta\rangle=30$ the average hard $p_{\rm T}$ is 16\% higher for events selected with V0M than that for events selected with flattenicity. The difference seems to increase at higher multiplicity densities. The results suggest that V0M and flattenicity are nearly equally sensitive to MPI, but flattenicity reduces the bias towards hard pp collisions. 

Figure~\ref{fig6} shows the angular correlations between the leading particle and associated particles. The leading particle is the one with the largest transverse momentum of the event. If $p_{\rm T}^{\rm trig}$ is the transverse momentum of the leading particle, then the associated particles are all those charged particles whose transverse momentum is lower than $p_{\rm T}^{\rm trig}$. In Fig.~\ref{fig6}, the charged-particle yield is reported as a function of $\Delta\varphi=\varphi-\varphi^{\rm trig}$ and $\Delta\eta=\eta-\eta^{\rm trig}$, where $\eta$ and $\varphi$ are the pseudorapidity and azimuthal angle of the associated particles, respectively. The left-hand (right-hand) side plot shows the angular correlation for the 0-1\% $1-\rho_{\rm nch}$ (V0M) class in pp collisions at $\sqrt{s}=13.6$\,TeV. According to table~\ref{tab:2}, the charged-particle multiplicity density at midpseudorapidity is around 26 for these event classes. While the selection based on V0M gives prominent jet structures at $\Delta\varphi=0$ and $\Delta\varphi=\pi/2$, for the pp sample selected with flattenicity, the near- and away-side peaks are significantly smaller than those observed when the selection is done in terms of V0M. This result is consistent with the isolation of more isotropic pp collisions in flattenicity classes than in V0M classes. Therefore, the event classifier is able to select high multiplicity pp collisions originated by several soft parton-parton scatterings producing a nearly homogeneous distribution of particles in both $\Delta\eta$ and $\Delta\varphi$. In PYTHIA~8 there are always $2\rightarrow2$ processes, therefore rather small near- and away-side peaks are observed. The size of the peaks are significantly smaller than those observed for pp collisions with similar charged-particle multiplicities but originated from harder processes. 

In summary, the sensitivity to MPI is still kept if multiplicity is used instead of average transverse momentum in the calculation of flattenicity. Moreover, with the actual detectors of the ALICE experiment, the analysis as a function of flattenicity seems to be feasible. In the next section, the light- and heavy-flavored hadron productions are studied for different $1-\rho_{\rm nch}$ event classes.

\section{\label{sec:results}Light- and heavy-flavor hadron production as a function of flattenicity}

\begin{figure*}[!ht]
\begin{center}
\includegraphics[width=0.46\textwidth]{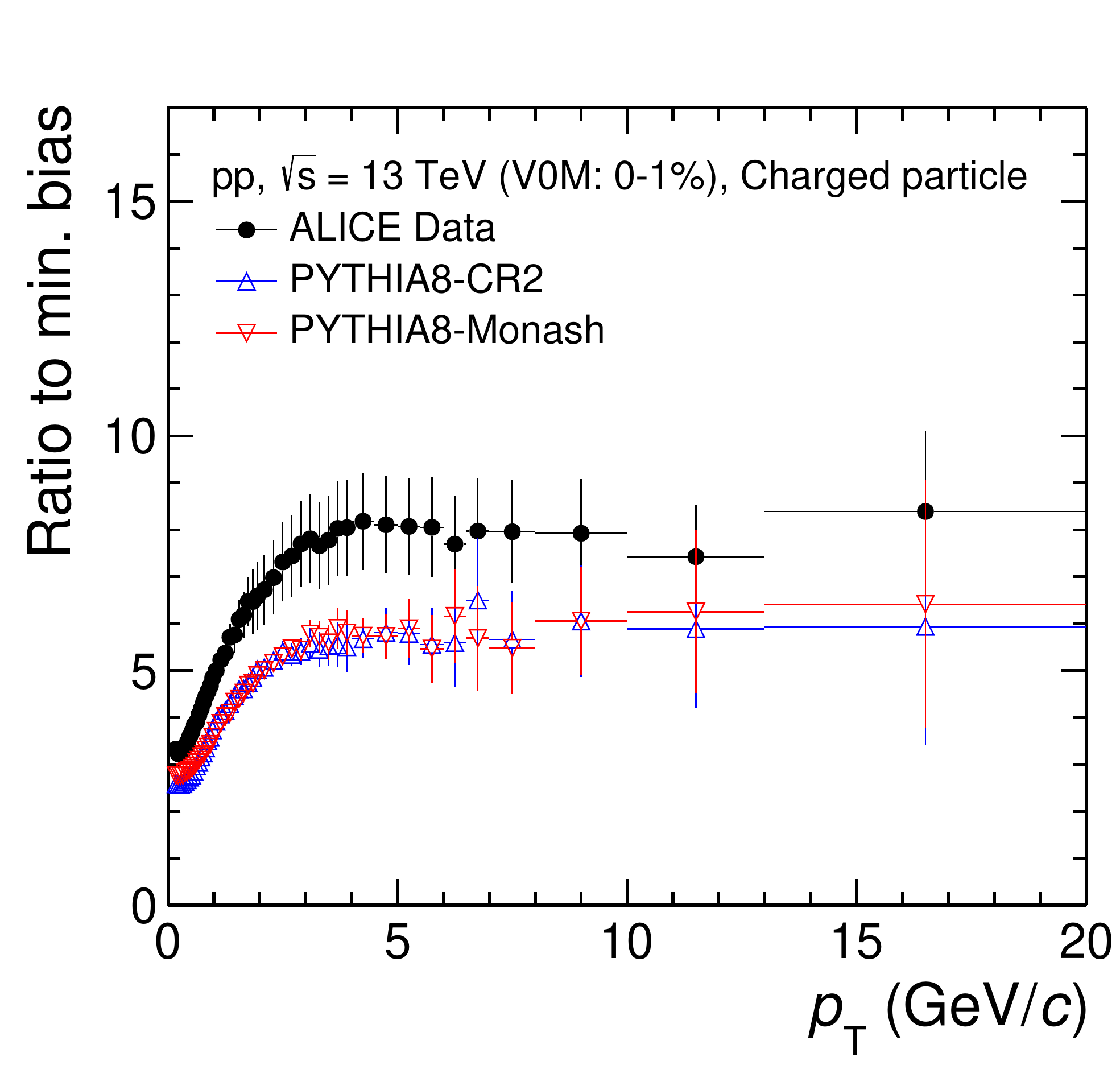}
\includegraphics[width=0.46\textwidth]{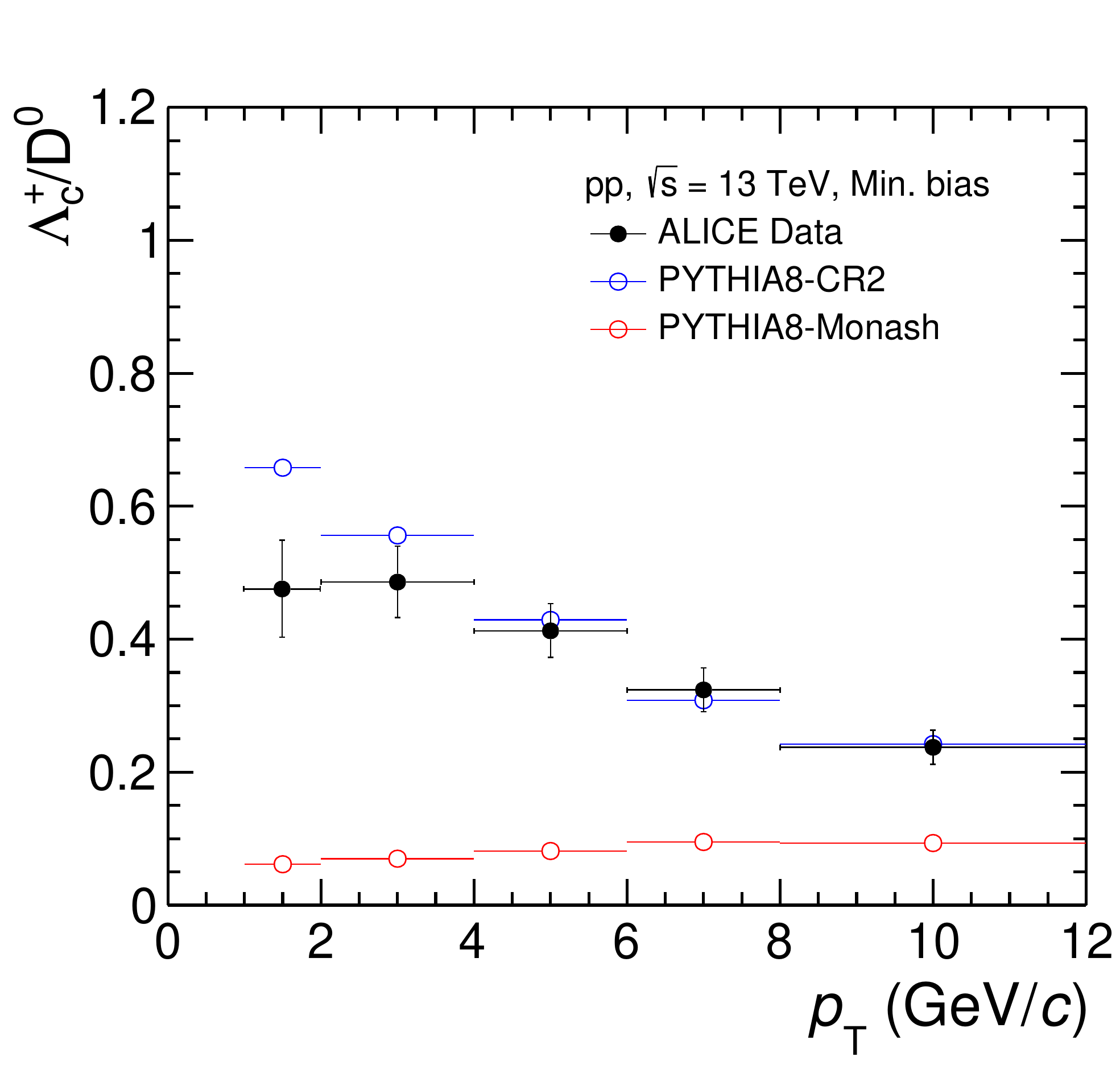}
\caption{The ratio of charged particle spectra in  $0-1\%$ V0M class to minimum bias in pp collisions with ALICE at $\sqrt{s}$ = 13 TeV are compared with the Pythia 8 Monash and CR2 predictions (left). The  $\Lambda_{\rm c}^{+}/{\rm D^{0}}$ ratios  in pp collisions with ALICE are compared to  Pythia 8 Monash and CR2 tunes (right).}
\label{results:DataComp}  
\end{center}
\end{figure*}

In this section, results using the color reconnection mode 2 of PYTHIA~8 is used to generate pp collisions at the LHC energies. This model is chosen given that it better describes the identified particle production than the Monash tune. The CR2 model includes junctions, which fragment into baryons, leading to an increased baryon production as compared to Monash tune. For example,  figure~\ref{results:DataComp} left panel shows the ratio between the $p_{\rm T}$ spectrum of charged particles in the 0-1\% \VZERO multiplicity class to the one measured in minimum bias pp collisions at $\sqrt{s}$ = 13 TeV by ALICE \cite{ALICE:2019dfi}. The data are compared with predictions from Pythia 8 Monash and color reconnection mode 2 (CR2). Both the models qualitatively reproduce the data very well, which shows the evolution of the spectral shape up to $\pt$ $\simeq$ 5 GeV/$c$ and the ratio is flat at high $\pt$. Figure~\ref{results:DataComp} right panel shows the measured $p_{\rm T}$-differential $\Lambda_{\rm c}^{+}/{\rm D^{0}}$ ratio in minimum bias collisions~\cite{ALICE:2021npz} along with model predictions. In particular, the CR2 model shows a very good agreement with the data.

\begin{figure}[!ht]
\begin{center}
\includegraphics[width=0.46\textwidth]{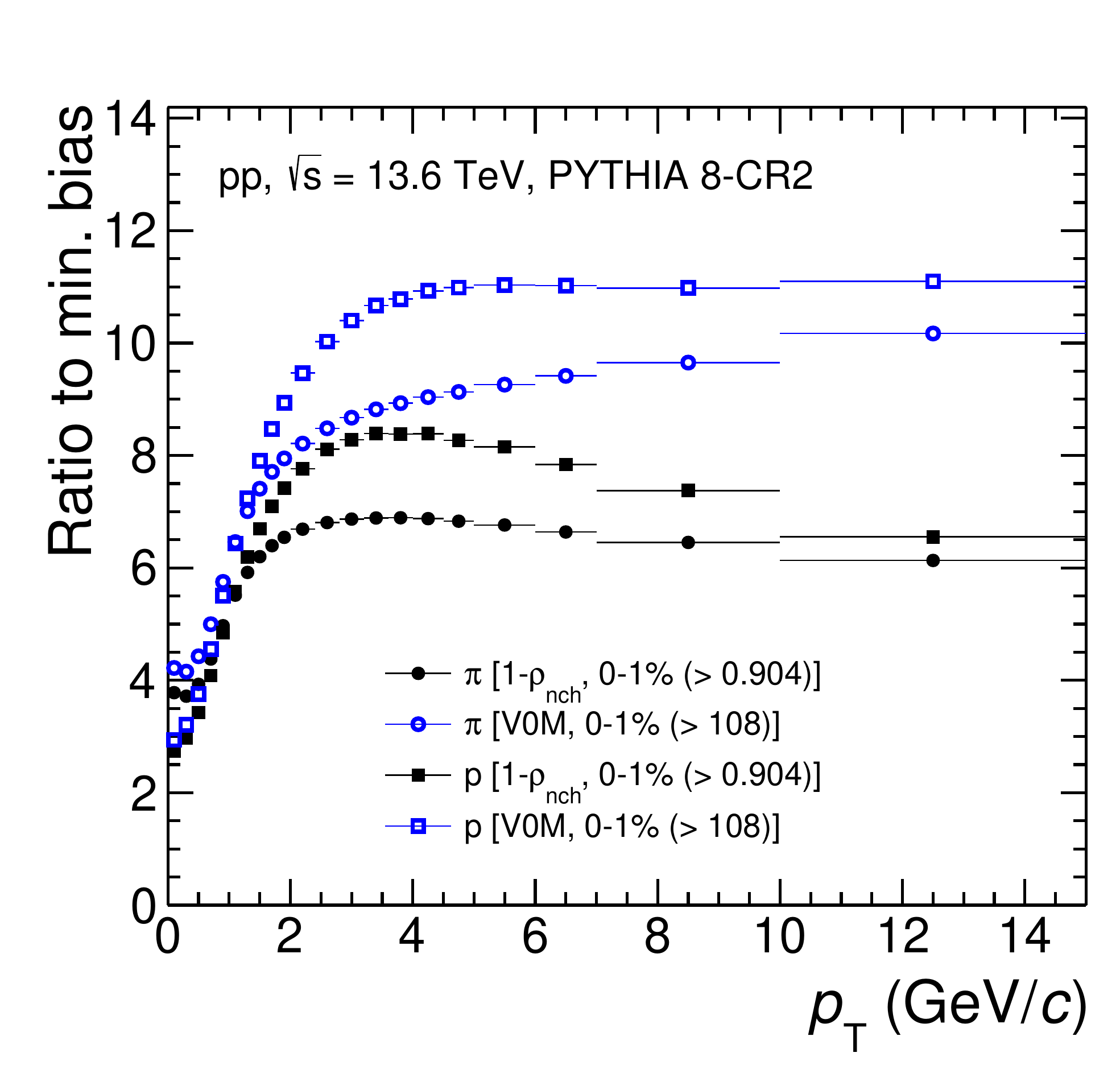}
\caption{Transverse momentum differential yield in the 0-1\% 1-$\rho_{\rm nch}$ (full markers) and 0-1\% V0M multiplicity (empty markers) class normalized to the yield in minimum-bias pp collisions. Results for pions and protons are displayed in round and square markers, respectively. The \pt spectra were calculated at mid-rapidity ($|y|<0.5$) in pp collisions at $\sqrt{s}=13.6$\,TeV simulated with PYTHIA~8.}
\label{results:fig1}  
\end{center}
\end{figure}

Figure~\ref{results:fig1} shows the charged-pion \pt spectrum in the 0-1\% V0M multiplicity class normalized to the charged-pion \pt spectrum in minimum-bias collisions.
\begin{figure*}[!ht]
\begin{center}
\includegraphics[width=0.96\textwidth]{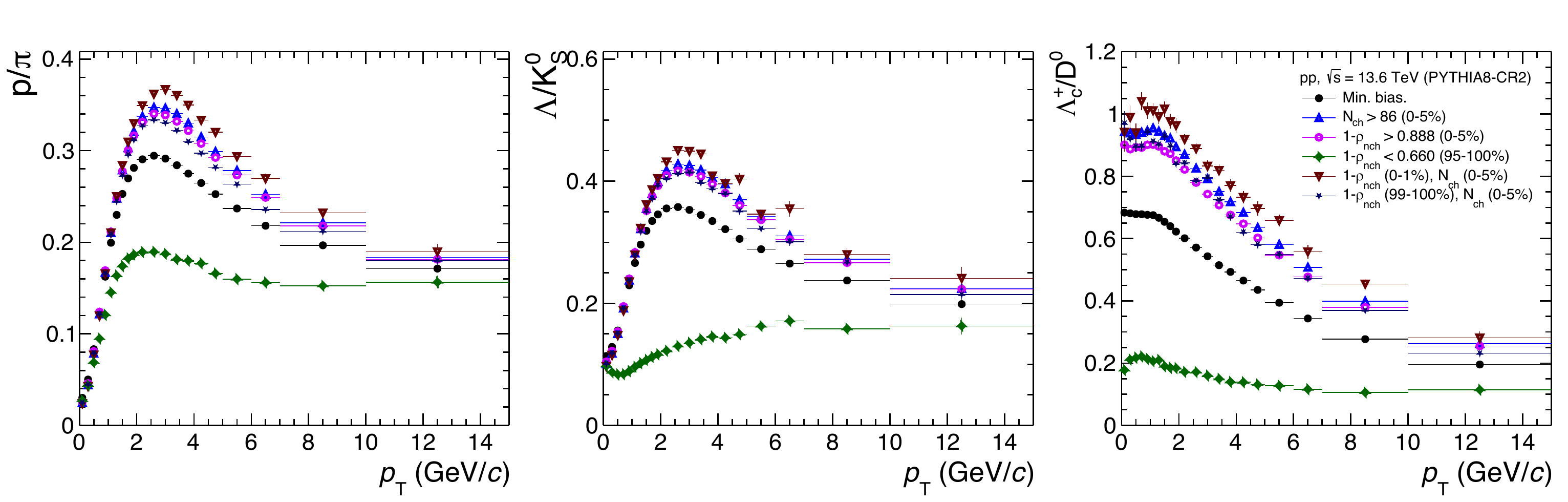}
\caption{$\pt$-differential baryon to meson ratios for particles containing light, strange and charm quarks, respectively. The selection of events is based on V0M multiplicity and 1-$\rho_{\rm nch}$ classes.}
\label{results:fig2}  
\end{center}
\end{figure*}
A comparison with the 0-1\% $1-\rho_{\rm nch}$ class is shown. While at low transverse momentum the ratios are very close to each other showing an increase up to $p_{\rm T}\approx2$\,GeV/$c$, at higher transverse momentum the ratios follow different trends. On one hand, the ratio for the 0-1\% V0M class exhibits a continuous rise with increasing \pt; on the other hand, the ratio for the 0-1\% $1-\rho_{\rm nch}$ class reaches a maximum that is followed by a reduction reaching a constant value of around 6. The ratios as a function of $1-\rho_{\rm nch}$ show a similar behavior as those reported as a function of the number of multi-partonic interactions~\cite{Ortiz:2020rwg}. The effect has been attributed to color reconnection, which according to Ref.~\cite{Ortiz:2013yxa} should originate a mass effect. The mass effect is tested using protons instead of pions. Figure~\ref{results:fig1} shows analogous ratios for protons. For similar event classes the effect gets significantly enhanced. Moreover, for the 0-1\% $1-\rho_{\rm nch}$ class a prominent bump structure is observed at intermediate transverse momentum. The effect is hidden in the 0-1\% V0M class given the presence of harder processes that produce a small increase of the ratio for $p_{\rm T}>6$\,GeV/$c$. The bump structure has not been observed in data since all the analyses have been performed so far as a function of the V0M multiplicity.

\begin{figure*}[!ht]
\begin{center}
\includegraphics[width=0.46\textwidth]{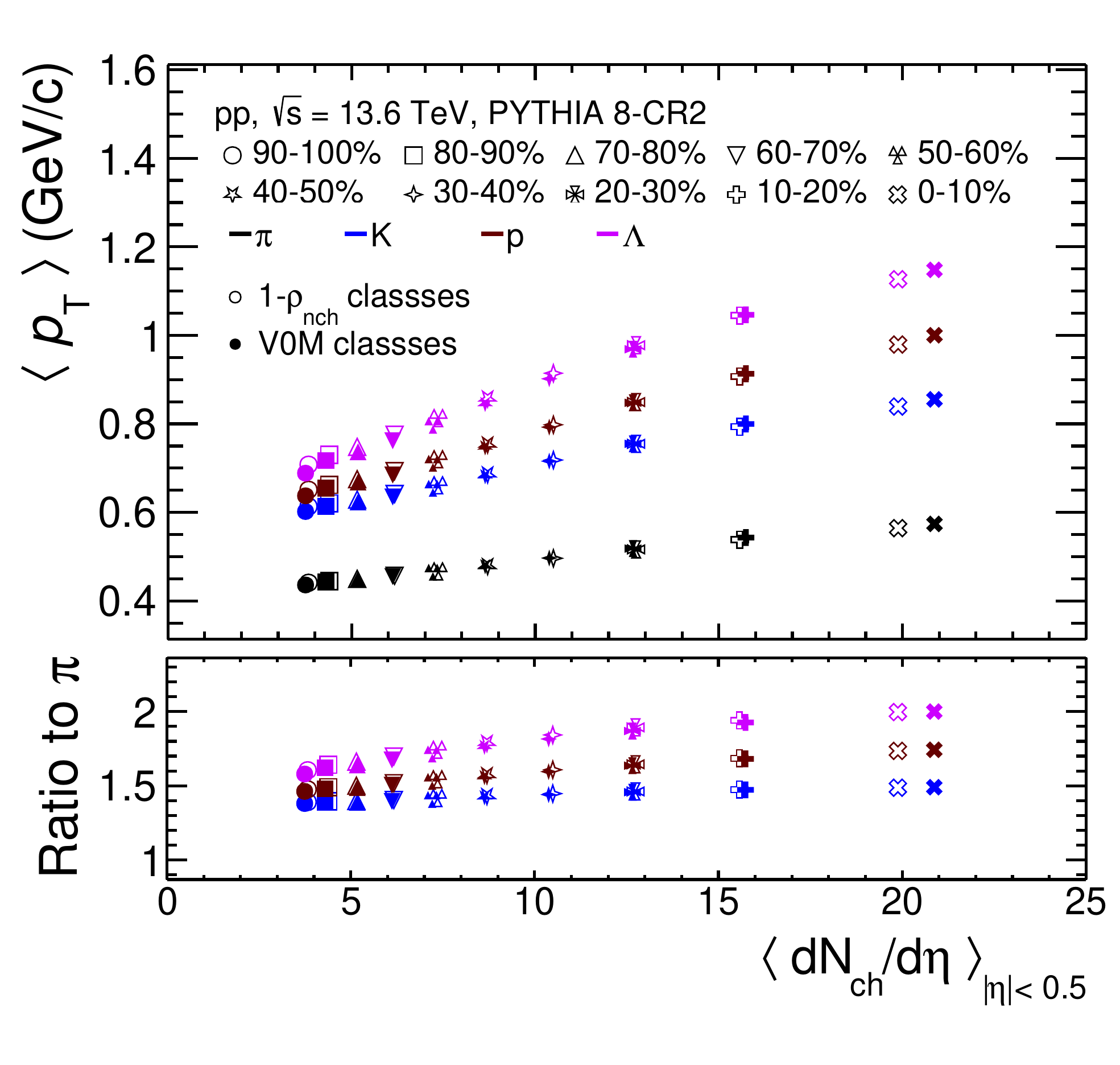}
\includegraphics[width=0.46\textwidth]{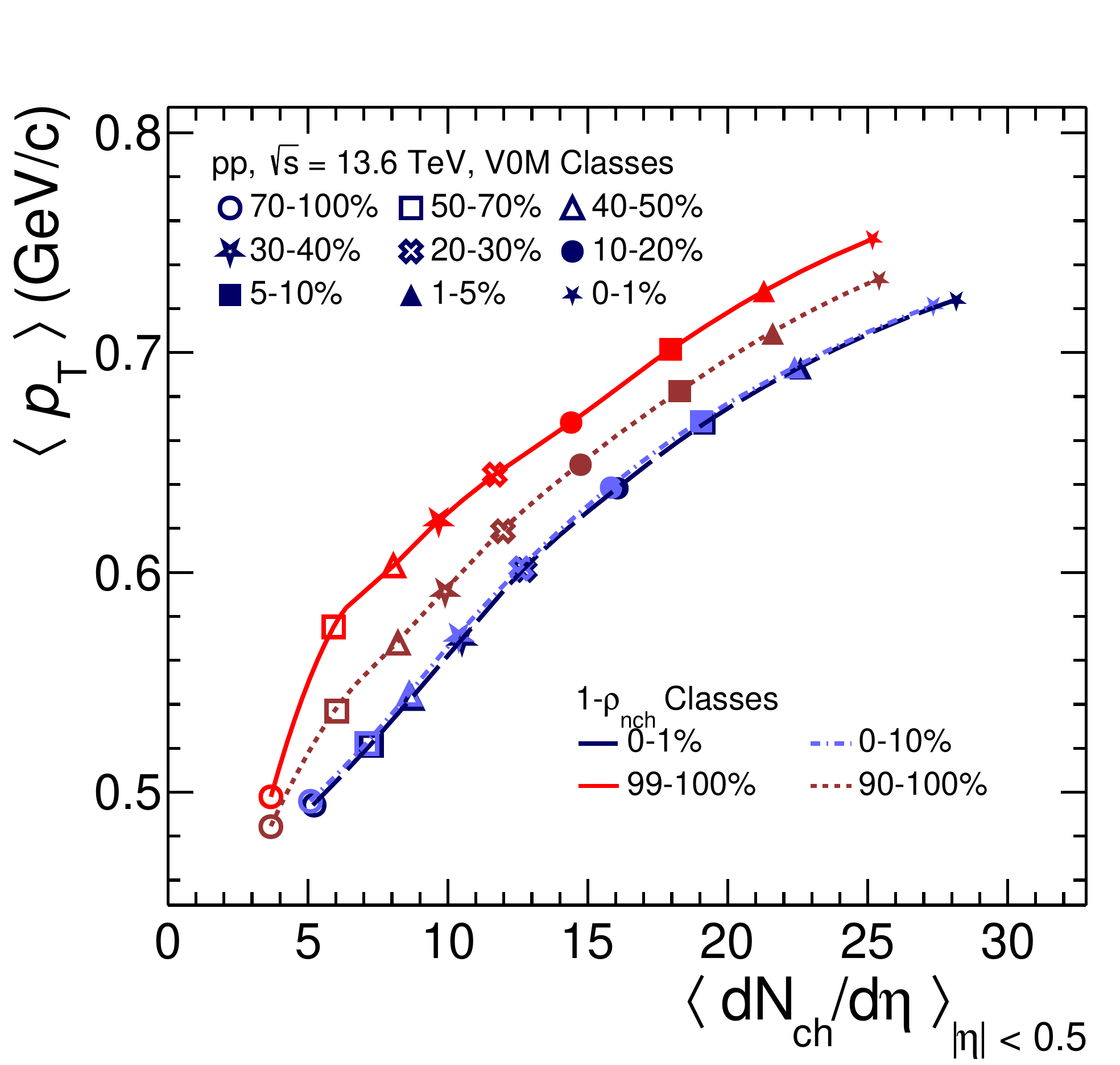}
\caption{Mean transverse momentum of identified particles calculated  in $|y|<0.8$ as a function of the charged-particle multiplicity at mid-rapidity (left). Here, we consider both V0M and 1-$\rho_{\rm nch}$ classes to see the effect of the mass-ordering  of $\langle p_{\rm T} \rangle$ on multiplicity and event-shape variable. Mean transverse momentum of the charged particles in V0M classes is shown as a function of the charged-particle multiplicity in the mid-rapidity, $|\eta| <$  0.5 for 1$\%$ and 10$\%$ top and bottom events in 1-$\rho_{\rm nch}$ classes (right).}
\label{results:meanPtVsNchIdentified}  
\end{center}
\end{figure*}

Last but not least, it has been reported that the \pt-differential baryon-to-meson ratios including $\Lambda_{\rm c}^{+}/{\rm D^{0}}$, ${\rm p}/\pi$ and $\Lambda/{\rm K_{s}^{0}}$ exhibit remarkable similarities. The ratios show a clear decrease  with increasing \pt in both pp and \pPb collisions in the range $2 < p_{\rm T} < 12$\,GeV/$c$.  At low \pt, predictions that include additional color-reconnection mechanisms beyond the leading-color approximation (CR2) describe rather well the overall features~\cite{ALICE:2020wfu}.  Figure \ref{results:fig2} shows the ${\rm p}/\pi$,  $\Lambda/{\rm K_{s}^{0}}$  and $\Lambda_{\rm c}^{+}/{\rm D^{0}}$ ratios as a function of the transverse momentum using V0M and 1-$\rho_{\rm nch}$ classes. All these ratios exhibit remarkably similar characteristics with a decreasing trend after \pt $\geq$ 2-3\,GeV/$\it{c}$. With the selection of 0-5$\%$ V0M or 1-$\rho_{\rm nch}$, we see an enhancement of these ratios at intermediate \pt that is around 18$\%$, 20$\%$, and 35$\%$ higher for ${\rm p}/\pi$,  $\Lambda/{\rm K_{s}^{0}}$  and $\Lambda_{\rm c}^{+}/{\rm D^{0}}$ as compared to minimum-bias pp collisions, respectively. One of the interesting observations is the shift of the peak structure towards higher momentum, which is often attributed to the radial flow effect for light-flavor particle production \cite{ALICE:2013wgn}.  The 0-1$\%$  1-$\rho_{\rm nch}$ class for the highest 0-5$\%$ V0M classes allows selecting events an isotropic event topology. From Fig.  \ref{results:fig2}, a clear picture is evolved where the baryon to meson ratio is further enhanced and for the first time we observe a clear peak structure for $\Lambda_{\rm c}^{+}/{\rm D^{0}}$ for pp collisions, which is earlier observed in \pPb collisions by ALICE collaboration\cite{ALICE:2020wfu}. This enhancement is suppressed for ${\rm p}/\pi$ ratio for the 95-100\% 1-$\rho_{\rm nch}$ event class that corresponds to jet topologies, whereas for  $\Lambda/{\rm K_{s}^{0}}$  and $\Lambda_{\rm c}^{+}/{\rm D^{0}}$ the peak structure at intermediate \pt is completely absent. The section of hard events (95-100\% 1-$\rho_{\rm nch}$) shows a similar trend for $\Lambda/{\rm K_{s}^{0}}$ ratio in jets as a function of \pt~\cite{Cui:2022gos}.  The selection of events based on charged-particle multiplicity or event topology should reflect on the $\langle p_{\rm T} \rangle$ of the identified particles. The left panel of Fig. \ref{results:meanPtVsNchIdentified} shows the $\langle p_{\rm T} \rangle$ of $\pi$, K, p and $\Lambda$ as a function of charged-particle multiplicity at mid-rapidly ($|y|<0.8$)  in the V0M and 1-$\rho_{\rm nch}$ event classes. A clear mass dependant evolution of $\langle p_{\rm T} \rangle$ is observed as a function of charged-particle multiplicity. However, we do not see a strong dependence on V0M or 1-$\rho_{\rm nch}$ event classes although we see harder spectra in the case of  V0M event classes for $\pi$ and p in Fig. \ref{results:fig1}. This is because the spectral shape doesn't significantly change at lower \pt, which is the dominant factor in the evaluation of the $\langle p_{\rm T} \rangle$. We further extend this study for charged particles by selecting 0-1\% and 0-10\% 1- $\rho_{\rm nch}$ event classes together with V0M selection. The $\langle p_{\rm T} \rangle$  of charged particles are quite similar for both subclasses and there is mild spread on the charged-particle multiplicity at mid-rapidity as seen from Fig.~\ref{results:meanPtVsNchIdentified}. However, by selecting 99-100\% and 90-100\% 1- $\rho_{\rm nch}$ event classes, a higher  $\langle p_{\rm T} \rangle$  is observed with the bottom 1$\%$, which is in agreement with the interpretation of selecting hard events.

\section{\label{sec:conclusions}Conclusions}

This paper presented a comprehensive study of flattenicity in pp collisions. The studies were performed using two tunes of the PYTHIA~8.307 Monte Carlo event generator. It has been demonstrated that the quantity is robust against variations in the size of the cell used for the calculation of flattenicity. Moreover, according to the studies, the event topology calculated in the pseudorapidity region covered by the ALICE \VZERO detector is strongly correlated with the event shape calculated in eight units of pseudorapidity. A comparison between the widely used V0M multiplicity estimator and flattenicity shows that although both of them show the same level of correlation with the MPI activity, flattenicity tends to select collisions involving several lower transverse momentum parton-parton scatterings than the V0M estimator. Moreover, the high multiplicity pp collisions selected with flattenicity yields a softer \pt spectrum than those isolated using the V0M multiplicity estimator. The results suggest that flattenicity can be used using the forward detectors that will operate during runs 3 and 4.

\begin{acknowledgments}
We acknowledge the useful comments of Andreas Morsch. Support for this work has been received from CONACyT under the Grants A1-S-22917. G. B. acknowledges the support received from the Hungarian National Research, Development and Innovation Office (NKFIH) under the contract numbers OTKA PD143226, OTKA FK131979, K135515, and the NKFIH grant 2019-2.1e.11-T\'ET-2019-00078. S.T. acknowledges the support under the INFN postdoctoral fellowship. A. O. acknowledges the support under the CERN Scientific Associateship and PASPA-UNAM. S. P. acknowledges the financial support from UGC, the Government of India

\end{acknowledgments}

\bibliography{flat}

\end{document}